\documentclass[14pt]{extarticle}
\usepackage[utf8]{inputenc}
\usepackage{mathtools}
\usepackage{amsfonts}
\usepackage{amsmath}
\usepackage{amssymb}
\usepackage{fancyhdr}
\usepackage{hyperref}
\usepackage{braket}
\usepackage{bbm}
\usepackage{faktor}
\usepackage{tensor}
\usepackage{cleveref}
\usepackage{nicefrac}

\newcommand\be{\begin{equation}}
\newcommand\ba{\begin{eqnarray}}
\newcommand\ee{\end{equation}}
\newcommand\ea{\end{eqnarray}}
\numberwithin{equation}{section}
\newcommand\bne{\begin{equation}}
\newcommand\ene{\end{equation}}
\newcommand\cH{\mathcal H}
\newcommand\del{\partial}

\usepackage[
   citestyle=numeric-comp,
   sorting=none,
   maxnames=2,
   maxbibnames=10,
   giveninits=true,
   mcite
  ]{biblatex}
 \renewbibmacro{in:}{}  
\AtEveryBibitem{\clearfield{month}}
\AtEveryBibitem{\clearfield{day}}
\AtEveryBibitem{\clearfield{doi}}

\DeclareFieldFormat{postnote}{#1}
\DeclareFieldFormat{multipostnote}{#1} 
\DeclareFieldFormat{pages}{#1}

\title{Entanglement distillation of boundary states of large $N$ $\mathrm{SU}(N)_1$, Chern-Simons theory and Riemann surfaces}

\author{Howard J. Schnitzer\footnote{schnitzr@brandeis.edu} \\ Department of Physics \\ Brandeis University \\ Waltham, MA 02454}

\begin{document}
\flushbottom
\maketitle

\thispagestyle{fancy}
\renewcommand{\headrulewidth}{0pt}

\begin{abstract}
A tree tensor network is proposed for the entanglement distillation of large $N$ $\mathrm{SU}(N)_1$ Chern-Simons theory and Riemann surfaces, adapting a proposal of Bao, et al. This is illustrated for the entanglement entropy $S(A)$ of a bipartite many-body system $A$, where here $S(A) = \log N$.
\end{abstract}

\clearpage

\section{Introduction}

There is a mapping $Z$, the functional integral, for Chern-Simons theory which relates the 3-dimensional manifold $M$ to a probability amplitude $Z(M)$. If $M$ has a boundary, the path integral selects a state $\ket{M}$ in a Hilbert space $\cH_{\del M}$ associated to the boundary field configurations. For $\mathrm{SU}(N)_1$ Chern-Simons theory one can construct generators of the Clifford group and stabilizer states, and compute the entanglement entropy $S(A)$ of a bipartite many-torus subsystem $A \subseteq \del M$ with only a single replica \cite{Schnitzer2019}. \Textcite{Salton2016} show that
\begin{align}
    S(A) & = - \log \frac{Z (-2M \cup_f 2M)}{Z(-M \cup_{\del M} M)^2} \label{eqn:1.1} \\
    & = \log N \label{eqn:1.2}
\end{align}
for $\mathrm{SU}(N)_1$ \cite{Schnitzer2019}.

Entanglement distillation, as presented by \mtextcite{BaoSet,*Bao:2018pvs,*Bao2019}, describes a state $\ket{\psi}$ in the boundary Hilbert space, which can be approximated by a state $\ket{\Psi}$ to arbitrary precision, by means of a tree tensor network relating the bulk and boundary theories. We explore this strategy for $\mathrm{SU}(N)_1$ Chern-Simons theory in this paper.

\section{Entanglement distillation}

Entanglement distillation, as discussed by \mtextcite{BaoSet}, presents a general procedure for constructing tensor networks for geometric states, which we adapt for a CFT/Chern-Simons theory correspondence. We review, and paraphrase, the essential features needed for our discussion.

Consider a state $\ket{\psi}$ in large $N$ $\mathrm{SU}(N)_1$ CFT, for a many-torus subsystem $A$, with the entanglement entropy of $\ket{\psi}$ between $A$ and its complement $A^c$. For $\mathrm{SU}(N)_1$ in the large $N$ limit, this is described by \eqref{eqn:1.1} and \eqref{eqn:1.2}. In addition to the von Neumann entropy, consider the Rényi entropies,
\begin{equation}
    S_\alpha(\rho) = \frac{1}{1-\alpha} \log \operatorname{Tr}\rho^\alpha
    \qquad
    (\alpha \geq 0)
\end{equation}
where $\rho$ is the density matrix of the subsystem $A$. $S_\alpha(\rho)$ is a monotonically decreasing function of $\alpha$, where
\begin{align}
    S_\text{max}(\rho) & = \lim_{\alpha \to 0} S_\alpha = \log[ \operatorname{rank} \rho], \\
    S_\text{min}(\rho) & = \lim_{\alpha\to \infty} S_\alpha = \log [\lambda^{-1}_\text{max}(\rho)].
\end{align}
Smooth min. and max. entropies satisfy \cite{Schnitzer2019, BaoSet}
\begin{align}
    S_\text{min} & = S - \mathcal{O}(\sqrt{S}) \\
    & = \log N - \mathcal{O}[(\log N)^\frac{1}{2}]
\end{align}
and 
\begin{align}
    S_\text{max} & = S + \mathcal{O}(\sqrt{S}) \\
    & = \log N + \mathcal{O}[(\log N)^\frac{1}{2}],
\end{align}
which is flat to leading order in large $N$. There exists a normalized state $\psi^\epsilon$ within $\epsilon$ trace distance of $\psi$ satisfying
\begin{align}
    \operatorname{rank}(\psi^\epsilon) & = N \exp [(\log N)^\frac{1}{2}], \\
    \lambda_\text{max}(\psi^\epsilon) & = \frac{1}{N} \exp[(\log N)^\frac{1}{2}].
\end{align}
The constraints on smooth min. and max. entropies for the multitorus Hilbert space permit entanglement distillation, where a bipartite state is approximated by a nearby state in which the entanglement between two regions is made manifest. The properties of smooth min. and max. entropies provide an approximate state
\begin{equation}
    \ket{\Psi} = \sum_{n=0}^{\exp \mathcal{O}(\sqrt{S})} \sum_{m=0}^{\exp[S-\mathcal{O}(\sqrt{S})]} \sqrt{\lambda_n} \ket{n,m}_A \ket{\bar{n},\bar{m}}_{A^c}
\end{equation}
where $(n,m)$ is an approximate division of eigenvalues into blocks
\begin{equation}
    n : \exp \mathcal{O} \sqrt{S} = \exp [(\log N)^\frac{1}{2}]
\end{equation}
of width
\begin{equation}
    m : \exp[S-\mathcal{O}(\sqrt{S})] = N \exp - [\mathcal{O}(\log N)^\frac{1}{2}]
\end{equation}
for $\mathrm{SU}(N)_1$. Then
\begin{align}
    |\braket{\Psi|\psi}|^2
    & \simeq 1 - \epsilon - \exp[-\mathcal{O}(\sqrt{S})] \nonumber \\
    & = 1- \epsilon- \exp[-\mathcal{O}(\log N)^\frac{1}{2}] \label{eqn:2.13}
\end{align}
where $\Psi$ approximates $\psi$ to the accuracy of \eqref{eqn:2.13}.

Introduce auxiliary Hilbert spaces to distill EPR states $\mathcal{H}_\gamma$ and $\mathcal{H}_f$ with dimensions
\begin{align}
    \dim \mathcal{H}_f & = \exp [\mathcal{O}(\log N)^\frac{1}{2}] \\
    \shortintertext{and}
    \dim \mathcal{H}_\gamma & = N \exp - [\mathcal{O}(\log N)^\frac{1}{2}]
\end{align}
which define maps $\mathcal{H}_f \otimes \mathcal{H}_\gamma \to \mathcal{H}_A$ and $\overline{\mathcal{H}}_f \otimes \overline{\mathcal{H}}_\gamma \to \mathcal{H}_{A^c}$, where the bond dimensions are $\dim \mathcal{H}_\gamma \gg \dim \mathcal{H}_f$ at large $N$. These maps are given by
\begin{equation}
    V \ket{n}_f \ket{m}_\gamma = \ket{n,m}_A
\end{equation}
and
\begin{equation}
    W \ket{\overline{n}}_{\overline{f}} \ket{\overline{m}}_{\overline{\gamma}} = \ket{\overline{n},\overline{m}}_{A^c}
\end{equation}
where $V$ and $W$ are isometries which embed Hilbert spaces of size
\begin{equation}
    \exp [ m \log N - \mathcal{O}(\sqrt{m})]
\end{equation}
and
\begin{equation}
    \exp[ \mathcal{O}(\sqrt{m})]
\end{equation}
and their complex conjugates into the physical space $\mathcal{H}_A \otimes \mathcal{H}_{A^c}$.

For a bipartite state of $\mathrm{SU}(N)_1$
\begin{equation} \label{eqn:2.20}
    \ket{\Psi} = (V \otimes W) (\ket{\phi} \otimes \ket{\sigma})
\end{equation}
with
\begin{equation}
    \ket{\phi} = \sum_{m=0}^{N \exp [ -\mathcal{O}(\log N)^\frac{1}{2}]} \ket{m \overline{m}}_{\gamma \overline{\gamma}}
\end{equation}
and
\begin{equation}
    \ket{\sigma} = \sum_{n=0}^{\exp [\mathcal{O}(\log N)^\frac{1}{2}]} \sqrt{\lambda_n} \ket{n \overline{n}}_{f \overline{f}}.
\end{equation}
Then $\ket{\Psi}$ is a tree tensor network of the form
\begin{equation}
    \Psi^{AA^c} = V^A_{f\gamma} W^{A^c}_{\overline{f}\overline{\gamma}} \phi^{\gamma \overline{\gamma}} \sigma^{f\overline{f}}
\end{equation}
with 
\begin{equation}
    |\gamma| = |\overline{\gamma}| = N \exp-\mathcal{O}[(\log N)^\frac{1}{2}]
\end{equation}
and
\begin{equation}
    |f|=|\overline{f}| = \exp \mathcal{O}[(\log N)^\frac{1}{2}].
\end{equation}
Identify
\begin{align}
    V^A_{f\gamma} & = N^A_{f\gamma} \\
    W^{A^c}_{\overline{f}\overline{\gamma}} & = N^{A^c}_{\overline{f}\overline{\gamma}}
\end{align}
where $N^A_{f\gamma}$ is the fusion matrix of $\mathrm{SU}(N)_1$ satisfying $A=f+\gamma \mod N$, where $f,\gamma$ and $A$ label the number of boxes of a single column Young tableau representation of $\mathrm{SU}(N)_1$, and where $\dim A = \dim \mathcal{H}_\gamma + \dim \mathcal{H}_f$. For the distilled state, to leading order in large $N$,
\begin{equation} \label{eqn:2.28}
    S(\Psi) = \log \dim A = \log N + \ldots.
\end{equation}
Therefore \eqref{eqn:2.20}--\eqref{eqn:2.28} represents a bipartite state on $\mathcal{H}_A \otimes \mathcal{H}_{A^c}$ to $\exp - \mathcal{O}[(\log N)^\frac{1}{2}]$ accuracy.

As an application we can discuss this in the context of a genus two Hilbert space $\mathcal{H}_{\Sigma_2}$ and a two-torus Hilbert space $\mathcal{H}_{T^2}^{\otimes 2}$. There is an isometry \parencite{Salton2016}
\begin{equation}
    X: \mathcal{H}_{T^2}^{\otimes 2} \to \mathcal{H}_{\Sigma_2}.
\end{equation}
If $U$ is a unitary operator on $\mathcal{H}_{T^2}^{\otimes 2}$, then
\begin{equation}
    Y = XUX^\dagger + (I-XX^\dagger)
\end{equation}
is unitary on $\mathcal{H}_{\Sigma_2}$. A representation of the genus two Riemann surface $\Sigma_2$ without boundaries is given by an octagon with identified sides labeled by
\begin{equation}
    aba^{-1}b^{-1}cdc^{-1}d^{-1}.
\end{equation}
Consider a bipartite division of this surface, where the isometry $X$ induces a bipartite division of $\mathcal{H}_{T^2}^{\otimes 2}$. [See Fig. 3 of ref. \cite{Salton2016}.] For the distilled state on $\mathcal{H}_{\Sigma_2}$, $U$ is approximated on $\mathcal{H}_{T^2}^{\otimes 2}$ to the same accuracy as $X^\dagger U X$ on $\mathcal{H}_{\Sigma_2}$.
There is an analogous discussion for $\mathcal{H}_{T^2}^{\otimes q} \to \mathcal{H}_{\Sigma_q}$, reflecting that \eqref{eqn:1.1} and \eqref{eqn:1.2} holds for a many-torus system.

\section{Concluding remark}

We proposed a tree tensor network for the entanglement distillation for large $N$ $\mathrm{SU}(N)_1$ Chern-Simons theory for a bipartite many-torus system. The next step in this program is the extension of the analysis to multipartite systems \cite{Nezami:2016zni,Bao:2015bfa,He:2019ttu}, in particular for stabilizer states. Note that equation \eqref{eqn:1.1} is reminiscent of the RT \cite{Ryu:2006bv} and HRT \cite{Hubeny:2007xt} relations for holography.

Other related applications of entanglement in Chern-Simons theory are discussed in \cite{Dong:2008ft,Balasubramanian:2016sro,Balasubramanian:2018por,Chun:2017hja,Dwivedi:2017rnj,Dwivedi:2019bzh,Pastawski:2015qua,Yang:2015uoa,Hayden:2016cfa,Hayden:2011ag,Balasubramanian:2014hda,GrossWalter2013}.

\section{Acknowledgements}

We thank Isaac Cohen, Alastair Grant-Stuart, and Andrew Rolph for their assistance in the preparation of the paper, which provides a contribution to the author's 85th birthday.

\printbibliography


\end{document}